\documentclass{article}
\usepackage{fullpage}
\usepackage{amsmath,amsfonts, graphicx}

\newtheorem{theorem}{Theorem}[section]
\newtheorem{lemma}[theorem]{Lemma}

\newcommand{\ds}{{\rm d}s}
\newcommand{\dt}{{\rm d}t}
\newcommand{\dr}{{\rm d}r}
\newcommand{\eps}{\epsilon}
\newcommand{\polylog}{{\rm polylog}}
\newcommand{\success}{{\rm success}}
\newcommand{\p}{\partial}
\newcommand{\smin}{{s_2^{\min}}}

\newcommand{\qed}{\hfill $\Box$ \medskip}

\newcommand{\ord}{{\tt A}}
\newcommand{\rand}{{\tt B}}

\begin{document}

\newcommand{\gl}{{\sc 3-gl}}
\newcommand{\ps}{p_{\rm success}}
\newcommand{\e}{{\rm e}}
\newcommand{\ex}[1]{{\rm E}[#1]}
\newcommand{\dg}{{\rm d}g}
\newcommand{\dz}{{\rm d}z}
\newcommand{\tmax}{t_{\rm max}}
\newcommand{\lmax}{\lambda_{\rm max}}

\title{How much backtracking does it take to color random graphs?\\
Rigorous results on heavy tails}

\author{\begin{tabular}[t]{c@{\extracolsep{1em}}c }
 Haixia Jia & Cristopher Moore\\ 
  hjia@cs.unm.edu  & moore@cs.unm.edu \\
 Computer Science Department &Computer Science Department\\
University of New Mexico & University of New Mexico  \\
Albuquerque NM 87131 & Albuquerque NM 87131
\end{tabular}} 

\maketitle

\begin{abstract} 
Many backtracking algorithms exhibit heavy-tailed distributions, in which their running time is often much longer than their median.  We analyze the behavior of two natural variants of the Davis-Putnam-Logemann-Loveland (DPLL) algorithm for Graph 3-Coloring on sparse random graphs $G(n,p=c/n)$.  Let $P_c(b)$ be the probability that DPLL backtracks $b$ times.   First, we calculate analytically the probability $P_c(0)$ that these algorithms find a 3-coloring with no backtracking at all, and show that it goes to zero faster than any analytic function as $c \to c^* = 3.847...$  Then we show that even in the ``easy'' phase $1 < c < c^*$ where $P_c(0) > 0$, including just above the emergence of the giant component, the expected number of backtracks is exponentially large with positive probability.  To our knowledge this is the first rigorous proof that the running time of a natural backtracking algorithm has a heavy tail for graph coloring.  
%Moreover, our results show that these algorithms take exponential time, not just below the 3-colorability threshold, but just above the degree $c=1$ at which the giant component first appears. 
In addition, we give experimental evidence and heuristic arguments that this tail takes the form $P_c(b) \sim b^{-1}$ up to an exponential cutoff.  
 \end{abstract}

\section{Introduction}

Many common search algorithms for combinatorial problems have been found experimentally to exhibit a heavy-tailed distribution in their running times; for instance, in the number of backtracks performed by Davis-Putnam-Logemann-Loveland (DPLL) algorithms on constraint satisfaction problems such as Satisfiability, Graph Coloring, and Quasigroup Completion~\cite{GentWalsh,GomesSelmanCrato,GomesSelmanKautz,HoggWilliams,SelmanKautzCohen}.  In such a distribution, with significant probability, the running time is much larger than its median, and indeed the expectation can be exponentially large even if the median is only polynomial.  These distributions typically take a power-law form, in which the probability that the algorithm backtracks $b$ times behaves as $P_c(b) \sim b^{-\gamma}$ for some exponent $\gamma$.  One consequence of this is that if a run of the algorithm has taken longer than expected, it is likely to take much longer still, and it would be a good idea to restart it (and follow a new random branch of the tree) rather than continuing to search in the same part of the search space.

\begin{figure}
\begin{center}
\includegraphics[width = 3in]{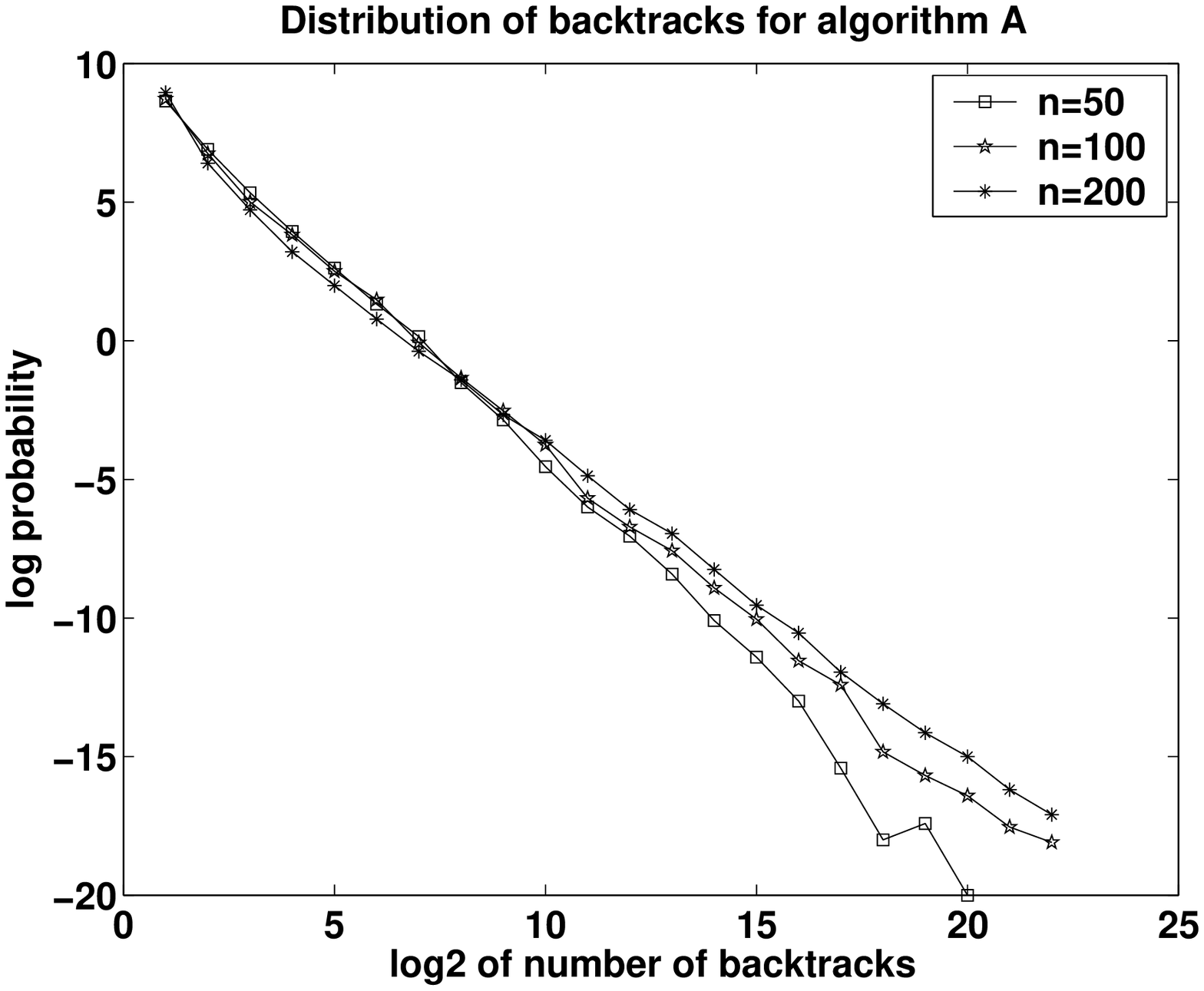}
\includegraphics[width =3in]{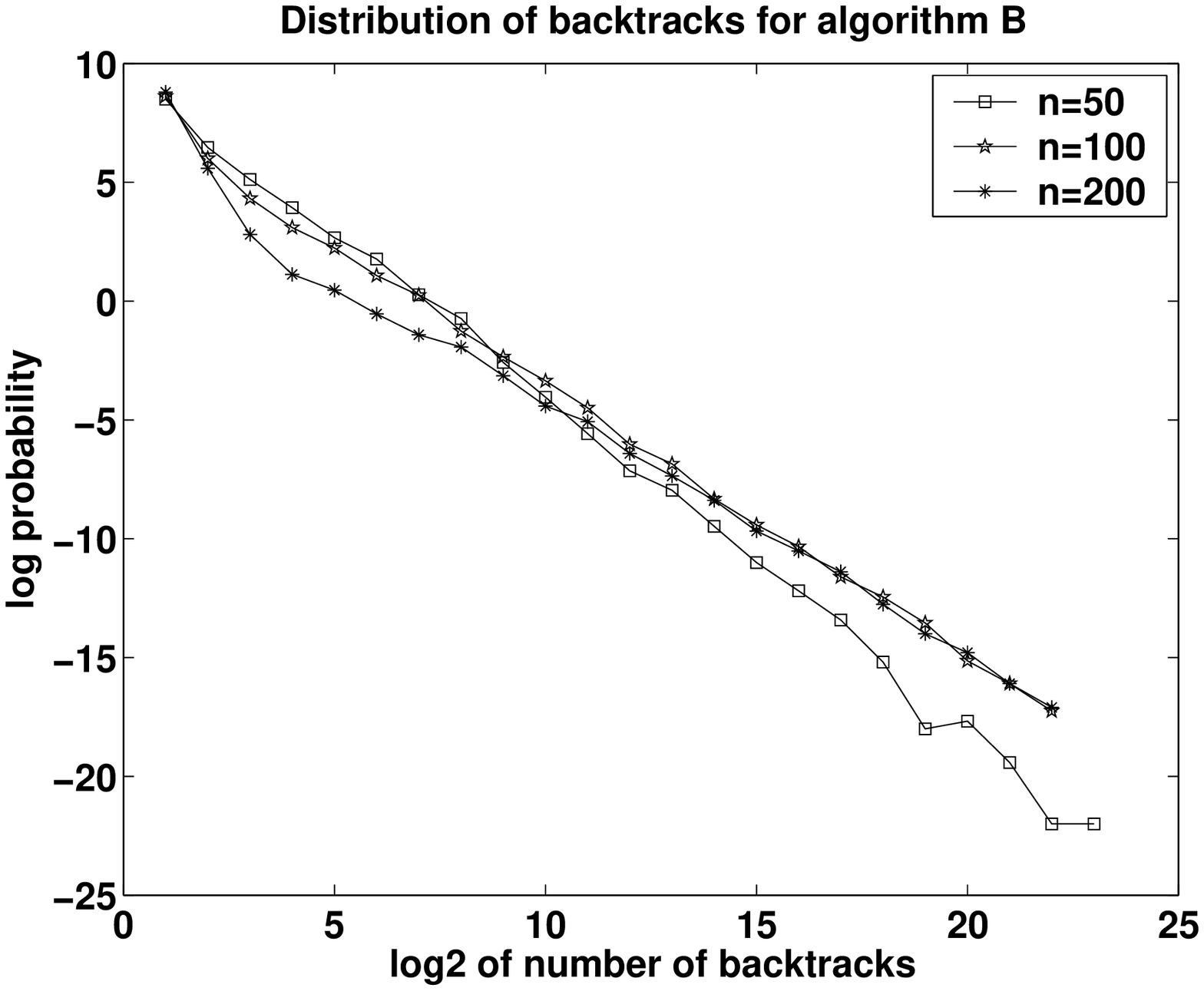}
\end{center}
\caption{Log-log plots of the distribution of the number of backtracks $P_c(b)$ for the two DPLL algorithms \ord\ and \rand\ described in the text on random graphs with $c = 3.5$.  The data appears to follow a power law $P_c(b) \sim b^{-1}$ in the limit $n \to \infty$.}
\label{fig:c35}
\end{figure}

For Graph 3-Coloring, in particular, these heavy tails were found experimentally by Hogg and Williams~\cite{HoggWilliams} and Davenport and Tsang~\cite{DavenportTsang}.  At first, it was thought that this heavy tail indicated that many {\em instances} are exceptionally hard.  A clearer picture emerged when Gomes, Selman and Crato~\cite{GomesSelmanCrato} found that the running times of randomized search algorithms on a typical {\em fixed} instance show a heavy tail.  In Figure~\ref{fig:c35} we show our own experimental data on the distribution of the number of backtracks for two versions of DPLL described below.  In both cases the log-log plot follows a straight line, indicating a power law.  As $n$ increases, the slopes appear to converge to $-1$, and we conjecture that $P_c(b) \sim b^{-1}$ up to some exponential cutoff.

A fair amount of theoretical work has been done on heavy tails, including optimal restart strategies~\cite{LubySinclairZuck} and formal models~\cite{ChenGomesSelman}.  However, there have been relatively few rigorous results establishing that these tails exist.  The most desirable result would be a proof, for some natural probability distribution over problems of size $n$, that $P_c(b) \sim b^{-\gamma}$ for some $\gamma$ in the limit of large $n$ and $b$.  To our knowledge, no such result has been obtained.  In this paper, we show a weaker result, namely that $b$ is exponentially large with positive probability, even for ``easy'' random problems where $b=0$ with positive probability (and, if $P_c(0) > 1/2$, the median value of $b$ is zero).  One related result is Achlioptas, Beame, and Molloy~\cite{AchBeameMolloy}, who showed using lower bounds on resolution proof complexity that DPLL takes exponential time on random instances of 3-SAT, even for some densities below the satisfiability threshold; our results appear to be the first on Graph Coloring, and we rely on much simpler reasoning.

Our results hold for two variants of DPLL.  Both of them are greedy, in the sense that they branch on a vertex with the smallest available number of colors; in particular, they perform {\em unit propagation}, in which any 1-color vertex is immediately assigned that color.  They are distinguished by which 2-color vertex they branch on when there are no 1-color vertices.  In algorithm \ord, the vertices are given a fixed uniformly random ordering, and we branch on the 2-color vertex of lowest index.  In algorithm \rand, we choose a vertex uniformly at random from among the 2-color vertices.   In both variants, we try the two possible colors of the chosen 2-color vertex in random order.  (How we branch on 3-color vertices is immaterial, since there is always a 1- or 2-color vertex while the algorithm is coloring the giant component.)

Our main result is the following:
\begin{theorem}
\label{thm:main}
For algorithms \ord\ and \rand, let $b$ be the number of times the algorithm backtracks on $G(n,c/n)$.  If $1 < c < c^* = 3.847...$, there exist constants $\beta, q > 0$ such that $\Pr[b > 2^{\beta n}] \ge q$, and so $\ex{b} = \Theta(2^{\beta n})$.
\end{theorem}
Although this theorem does not show that the tail of $P_c(b)$ behaves as $b^{-1}$, we believe our arguments can be refined to do that.  Along the way, we calculate the precise probability that these algorithms succeed with no backtracking at all:
\begin{theorem}
\label{thm:p0} 
Let $1 < c < c^* = 3.847...$  For algorithms \ord\ and \rand, the probability the algorithm colors $G(n,c/n)$ without backtracking is 
\begin{equation}
\label{eq:p0t}
 P_c(0) = \exp\left( - \int_0^{t_0} \dt \,\frac{c \lambda^2}{2 (1-\lambda)(2+\lambda)} \right) + o(1) 
\end{equation}
where $\lambda = (2/3)c(1-t-\e^{-ct})$ and $t_0$ is the smallest positive root of $1 - t - \e^{-ct} = 0$.
\end{theorem}
We note below that $P_c(0)$ approaches zero faster than any analytic function as $c$ approaches $c^*$, and comment on the fact that this ``essential singularity'' makes it very difficult to locate the threshold at which such heuristics succeed using numerical experiments.  

Our work is motivated partly by recent results of Ein-Dor and Monasson~\cite{MonassonColoring}.  Suppose the expected amount of backtracking takes the form $\exp(\omega(c)n + o(n))$; then, based on an earlier analysis of 3-SAT by Cocco and Monasson~\cite{MonassonSAT}, they estimate $\omega(c)$ by modeling the search tree with a time-dependent branching process.  The values of $\omega(c)$ they obtain using this approach agree very well with experiment, especially when the average degree is large.  Beame, Culberson, Mitchell and Moore~\cite{beame} proved for some DPLL algorithms that $\omega(c) = O(1/c^2)$ in the limit of large $c$, in agreement with a scaling argument of~\cite{MonassonColoring}.  However, their arguments do not apply as well for small values of $c$,  below the 3-colorability threshold.

The idea behind Theorem~\ref{thm:main} is very simple.  Partway down a random branch of the tree, with positive probability, the subgraph induced by the remaining vertices contains a small subgraph, which is not list-colorable given its remaining colors; say, a triangle composed of the vertices whose available colors are red and green.  No matter what the algorithm does from that point on, it will encounter this subgraph over and over again, vainly recoloring other vertices in the hope that it will go away.  Thus every branch of this subtree will fail, and the algorithm is forced to backtrack to before this subgraph's neighbors were colored.  The result is that there is a strong positive correlation between the events that two different branches of the search tree fail, and so an exponentially large number of branches can fail even though a given one succeeds with positive probability.  

We will rely heavily on the fact that for both these variants of DPLL, a single random branch is equivalent to a linear-time greedy heuristic, \gl, analyzed by Achlioptas and Molloy~\cite{AchMolloy}.  They showed that if $1 < c < c^*$ where $c^* = 3.847...$ then \gl\ colors $G(n,c/n)$ with positive probability.  (If $c<1$ then the graph with high probability has no bicyclic component and \gl\ colors it with probablity $1$.)  This shows that $P_c(0) > 0$, i.e., with positive probability these variants of DPLL succeed with no backtracking at all.  However, as our results show, the expected amount of backtracking is exponentially large even for random graphs with $c$ in this ``easy'' regime, and indeed just above the appearance of the giant component at $c = 1$.

The paper is organized as follows.  In Section~\ref{sec:p0}, we prove Theorem~\ref{thm:p0} by looking closely at \gl\ using the techniques of Achlioptas and Moore~\cite{AchMoore}, grouping the steps of the algorithm into rounds, and exactly analyzing the correlations between the 1-color vertices colored in a given round.  We also use generating functions to calculate the distribution of the number of 1-color vertices at a given time.  

In Section~\ref{sec:exp} we prove Theorem~\ref{thm:main} along the lines alluded to above.  First we show that a triangle of red-green vertices appears with positive probability, dooming an entire subtree; then, we show that for both variants of DPLL, with positive probability the number of leaves of this subtree is exponentially large.  Finally, in Section~\ref{sec:dis} we conclude and give some intuition about how Theorem~\ref{thm:main} might be strengthened to prove that the number of backtracks is distributed as a power law.

%Finally, in Section~\ref{sec:powerlaw} we give experimental evidence that the number of backtracks has a power-law tail, $P_c(b) \sim b^{-1}$, and offer some intuition for why this might be so.

%While we focus on $c < 3.847...$, this type of positive correlation should exist for larger values of $c$ as well (although Ein-Dor and Monasson's results suggest that it becomes less important as $c$ grows).  This suggests a rationale for random restarts: namely, that if multiple branches in a particular subtree fail, it may be due to a small un-list-colorable subgraph created near the root of that subtree, and so trying another branch within that same subtree may be less likely to find a solution than starting from the beginning and trying a new branch entirely.  This also helps justify the approach of ``subgraph caching'' in which we try to identify these subgraphs early on, rather  than banging our heads against them an exponential number of times.
% talk with Culberson about this

%Finally, we comment on some preliminary efforts to remove the heavy tail in the number of backtracks, by ``looking ahead'' and trying to identify these uncolorable subgraphs early on.  

We use red, green, and blue to denote our three colors.  All asymptotics are in the limit of large $n$, and we omit floors and ceilings.

\section{The probability of success without backtracking}
\label{sec:p0}

\subsection{\gl\ and differential equations}

Achlioptas and Molloy~\cite{AchMolloy} analyzed a greedy list-coloring heuristic they call \gl.  Each vertex $v$ has a list $\ell(v)$ of available colors, which are removed when they are assigned to its neighbors.  We call $v$ a {\em $q$-color vertex} if $|\ell(v)| = q$ and every vertex is 3-color
vertex at the beginning.  Then \gl\ works as follows:
\begin{enumerate}
\item If there are any 1-color vertices, choose one at random and assign its available color to it.
\item Else if there are 2-color vertices, choose one $v$ at random, and assign it a random color $c \in \ell(v)$.
\item Else choose a 3-color vertex at random and assign a random color to it.
\end{enumerate}
Everything outside the giant component of $G(n,c/n)$ with high probability consists of trees and unicyclic components, and it is easy to see that \gl\ succeeds on such components.  Therefore, we focus on the phase of \gl\ which colors the giant component, during which there is always a 1- or 2-color vertex.  We refer to steps of type (1) and (2) above, in which we color 1-color and 2-color vertices, as ``forced'' and ``free'' respectively.  It will be useful to follow Achlioptas and Moore~\cite{AchMoore} and group steps into ``rounds,'' where each round consists of a free step followed by a cascade of forced steps.  

Since the first branch of both our variants of DPLL is equivalent to a run of \gl, the probability $P_c(0)$ that they color the graph with no backtracking at all is the same as the probability that \gl\ succeeds, i.e., that it colors the entire graph without creating a $0$-color vertex.  This in turn is the probability that all of \gl's rounds succeed.

Now, define the {\em state} of a round as the number of uncolored vertices of each color list, i.e., the number of 3-color vertices and the number of 2-color vertices of each color pair, present at the beginning of that round (by definition there are no 1-color vertices present).  By the principle of deferred decisions, the uncolored part of the graph is uniformly random in $G(n',p)$ where $n'$ is the total number of uncolored vertices.  Therefore, the probability that a given round fails is a function only of its state.  Moreover, if we condition on the state of each round, the events that various rounds fail become independent, and $P_c(0)$ is simply the product over all rounds of the probability that they succeed.  

As it turns out, the probability that a given round succeeds is a continuous function of its state, so to calculate $P_c(0)$ within $o(1)$ it is sufficient to estimate the state to within $o(n)$.  The technique of differential equations, and in particular Wormald's theorem~\cite{Wormald}, allows us to do this.  Let $S_2(R)$ and $S_3(R)$ be the number of 2- and 3-color vertices at the beginning of the $R$'th round.  Then, the behavior of \gl\ on $G(n,c/n)$ can be modeled with the following set of differential equations in the ``rescaled'' variables $s_3$ and $s_2$, where the variable of integration is $r=R/n$~\cite{AchMolloy,AchMoore}:
\begin{eqnarray}
\frac{\ds_3}{\dr} & = & - \frac{c s_3}{1-\lambda} , \quad s_3(0) = 1 \nonumber \\
\frac{\ds_2}{\dr} & = & \frac{c s_3  - 1}{1-\lambda}, \quad s_2(0) = 0 
\label{eq:diffeq}
\end{eqnarray}
where
\[ \lambda = \frac{2}{3} c s_2 \enspace . \]
Specifically, let $s_3(r)$ and $s_2(r)$ be the solutions to~\eqref{eq:diffeq}, and let $r_0$ be the smallest positive root of $s_2(r) = 0$.  Then the following event holds with high probability: $S_3(R) = s_3(R/n) n + o(n)$ and $S_2(R) = s_2(R/n) n + o(n)$, with $s_2(R/n) n/3 + o(n)$ 2-color vertices of each color pair, uniformly for all $R$ with $0 < R/n < r_0$.  Since Achlioptas and Molloy~\cite{AchMolloy} showed that \gl\ succeeds with positive probability, this event holds with high probability even when we condition on the event that \gl\ succeeds.

We briefly review how the differential equations~\eqref{eq:diffeq} are derived.  The idea is that each round can be modeled by a branching process in which coloring a vertex $v$ causes some of $v$'s 2-color neighbors to become 1-color vertices.  {\em A priori} we have a 3-type branching process, consisting of the 1-color vertices of the three colors, with a $3 \times 3$ transition matrix $M$ whose entries depend on the number of 2-color vertices of the three color pairs; for instance, the expected number of red 1-color vertices created by coloring a vertex blue is $p$ times the number of red-blue vertices.  This results in a system of four coupled differential equations, which we omit here.  However, since both this system and the initial conditions are symmetric under permutations of the colors, its trajectory is symmetric as well, and we can reduce it to the smaller system~\eqref{eq:diffeq}.  In that case there are with high probability $s_2 n/3 + o(n)$ 2-color vertices of each color pair, so we have
\begin{equation} \label{eq:m}
M = \frac{c}{3} \left( \begin{array}{ccc} 
0 & s_2 & s_2 \\ s_2 & 0 & s_2 \\ s_2 & s_2 & 0 \
\end{array} \right)
= \frac{1}{2} 
\left( \begin{array}{ccc} 
0 & \lambda & \lambda \\ \lambda & 0 & \lambda \\ \lambda & \lambda & 0
\end{array} \right) \enspace . 
\end{equation}
and $M$'s only nonzero eigenvalue is $\lambda$, the total expected number of 1-color vertices created per step.

If $\lambda < 1$ this branching process is subcritical, and the expected number of initially 2-color vertices colored during a round is $1/(1-\lambda) - o(1)$.  Here $o(1)$ includes the probability that the graph induced by these vertices is not a tree (including the probability that a 0-color vertex is created and the round fails).  These vertices have an expected number $p S_3/(1-\lambda) - o(1) $ of 3-color neighbors; only $o(1)$ of these become 1- or 0-color vertices, and the rest become 2-color vertices.  Rescaling according to Wormald's theorem then yields the differential equations~\eqref{eq:diffeq}.  

To solve~\eqref{eq:diffeq}, it is convenient to change the variable of integration from $r$ to $t$, where $T=tn$ is the number of steps (free and forced) taken so far.  Using $\dt/\dr = 1/(1-\lambda)$, this gives the original differential equations derived in~\cite{AchMolloy}:
\begin{eqnarray}
\frac{\ds_3}{\dt} & = & - c s_3, \quad s_3(0) = 1 \nonumber \\
\frac{\ds_2}{\dt} & = & c s_3  - 1, \quad s_2(0) = 0 
\label{eq:diffeqt}
\end{eqnarray}
The solution to~\eqref{eq:diffeqt} is easily seen to be
\begin{equation}
\label{eq:sol}
 s_3(t) = \e^{-ct}, \quad s_2(t) = 1 - t - \e^{-ct} 
\end{equation}
Maximizing $s_2(t)$ shows that $\lambda < 1$ for all $t$ if and only if $c < c^*$ where $c^* = 3.847...$ is the smallest positive root of $c - \ln c = 5/2$.  Using the $-1$st branch of Lambert's function, defined as $W(x)=y$ where $y=x \e^x$, we can write $c^* = -W_{-1}(-\e^{-5/2})$.

The number of rounds performed after $T$ steps is with high probability $r(T/n) n + o(n)$, where
\begin{equation}
\label{eq:r}
 r(t) = \int_0^t \dt \,(1-\lambda) 
  = \frac{2}{3} \left(1-\e^{-c t}\right) 
  + \left(1 - \frac{2c}{3} \right)t
  + \frac{c}{3} t^2 \enspace . 
 \end{equation}
We will use this in the proof of Theorem~\ref{thm:main} below.

\subsection{Proof of Theorem~\ref{thm:p0}}

In this section we use the branching process associated with \gl\ to calculate the probability that a given round succeeds.  As we argued above, conditioning on the state at the beginning of each round makes the events that they succeed independent.  Taking the product of these probabilities then gives~\eqref{eq:p0t} and proves Theorem~\ref{thm:p0}.

\begin{lemma}
\label{lem:round}
Suppose that the state of a round $R$ contains $s_2 n/3 + o(n)$ 2-color vertices of each color pair, where $\lambda \equiv (2/3) c s_2 < 1$.  Then the probability that $R$ succeeds is
\begin{equation} \label{eq:qsuccess}
q_\success(r) = 1 - \frac{f(\lambda)}{n} + o(1/n) 
%= \exp\left( - \frac{f(\lambda(r))}{n} \right) + o(1/n)
\end{equation}
where
\begin{equation}
\label{eq:f}
f(\lambda) = \frac{c \lambda^2}{2 (1-\lambda)^2 (2+\lambda)} 
\enspace .
\end{equation}
\end{lemma}

\noindent
{\bf Proof.}
% of Lemma~\ref{lem:round}.}
We associate $R$ with a tree $T$ as follows: let $T$'s edges consist of the pairs $u,v$ such that coloring $u$ removes a color from $\ell(v)$.  Then $T$ spans the subgraph induced by the vertices colored, plus any 0-color vertices created, during $R$.  We will say that $R$ {\em generates} $T$.  

Now, the probability that $R$ fails is clearly a function of the tree it generates, and the probability it generates a given tree is a function only of its state.  Since $\lambda < 1$, the branching process corresponding to $R$ is subcritical, and arguments analogous to~\cite{AchMoore} show that the probability that $R$ generates a given tree differs by $o(1)$ from the probability that the branching process generates a tree of the same type.  This probability in turn is a continuous function of the entries of its transition matrix, and therefore of~$\lambda$.  Finally, since the size $t$ of the tree generated by a subcritical branching process has an exponential tail, its second moment $\ex{t^2}$ is finite; since $R$ fails with probability at most $p t^2 = O(t^2/n)$, averaging over all trees gives a probability of failure $\Theta(1/n)$, justifying the scaling inherent in~\eqref{eq:qsuccess}.

We now calculate $f(\lambda)$, i.e., $n$ times the probability that a round fails, within the branching process model.  First, suppose a round starts (on its free step) by coloring a vertex red.  Then, using~\eqref{eq:m}, the expected number of 1-color vertices of each color generated by the round is~\cite{AchMoore}
\[ (1-M)^{-1} \cdot \left( \begin{array}{c} 1 \\ 0 \\ 0 \end{array} \right)
= \frac{1}{(1-\lambda)(2+\lambda)} 
\left( \begin{array}{c} 2 - \lambda \\ \lambda \\ \lambda \end{array} \right)
\]
i.e., $(2-\lambda)/((1-\lambda)(2+\lambda))$ red vertices (including the initial one) and
$\lambda/((1-\lambda)(2+\lambda))$ each of the other two colors.  (Note that the total expected number of vertices is $1/(1-\lambda)$, but their colors are correlated with the color of the initial vertex.)

Now, it is not the case that the probability of failure in a round is $p$ times the number of pairs of 1-color vertices with the same color in $T$.  For instance, if a red 1-color vertex $u$ is colored before $v$ becomes a red 1-color vertex, $u$ and $v$ cannot be connected, since if they were $v$ would have become a 1-color vertex (of a different color) when we colored $u$.  The only ``dangerous'' pairs where coloring $u$ might make $v$ a $0$-color vertex are those where $v$ is present, but not yet colored, when we color $u$.

\begin{figure}[ht]
\centerline{\includegraphics[height=2in]{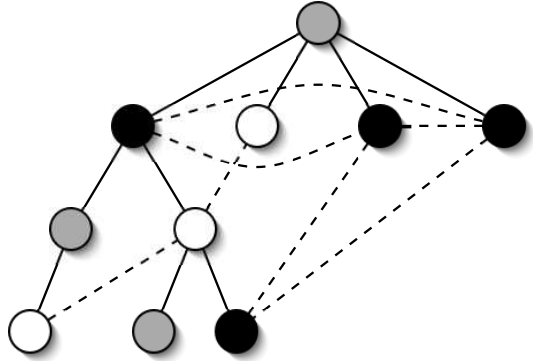}}
\caption{The dangerous pairs in a round.  The root is the vertex chosen on the free step, and siblings are ordered with the youngest on the left.}
\label{fig:uncles}
\end{figure}

It is easy to see that whether or not a round fails does not depend on the order in which we color the $1$-color vertices (although which vertex becomes a 0-color vertex does).  Therefore, although \gl\ chooses from the $1$-color vertices randomly, we can assume instead that we always color the youngest 1-color vertex, and thus perform a depth-first traversal of the tree $T$.  A little reflection shows that the dangerous pairs are then those $u,v$ where $v$ is an older sibling of $u$, an ``uncle'' which is older than $u$'s parent, or a great-uncle older than $u$'s grandparent, and so on.  In Figure~\ref{fig:uncles} we show part of a round, and connect the dangerous pairs with dotted lines.  If there are $D$ such pairs, the expected probability that the round succeeds is then $\ex{(1-p)^D} = 1-c \ex{D}/n+o(1)$, so $f(\lambda) = c \ex{D}$.

Each vertex $v$ in the branching process has a number of children $m$ which is Poisson-distributed with mean $\lambda$, and the number of dangerous pairs below $v$ includes those below each of its children.  In addition, if $v$ is red, its children are green and blue; given a pair of siblings $x$ and $y$  where $x$ is younger, the number of additional dangerous pairs is either the number of green descendants at or below $x$ or the number of blue ones, depending on $y$'s color.  Since the expected number of green or blue descendants at or below a green or blue vertex is $2/((1-\lambda)(2+\lambda))$, $y$ takes each of these colors with probability $1/2$, and the expected number of pairs of siblings is $\ex{{m \choose 2}} = \lambda^2/2$, we have
\[ \ex{D} = \lambda \ex{D} + \frac{1}{2} \frac{\lambda^2}{2} \frac{2}{(1-\lambda)(2+\lambda)} \] 
and so
\[ \ex{D} = \frac{\lambda^2}{2 (1-\lambda)^2 (2+\lambda)} \] 
Setting $f(\lambda) = c \ex{D}$ gives~\eqref{eq:f} and completes the proof.
\qed 

Lemma~\ref{lem:round} then implies the following.

\begin{lemma}
\label{lem:p0} 
Let $1 < c < c^* = 3.847...$  The probability that \gl\ succeeds on $G(n,c/n)$, and that algorithms \ord\ and \rand\ color $G(n,c/n)$ without backtracking, is 
\begin{equation}
\label{eq:p0}
 P_c(0) = \exp\left(- \int_0^{r_0} \dr f(\lambda(r)) \right) + o(1) 
\end{equation}
where $f(\lambda)$ is given by~\eqref{eq:f}, $\lambda(r) = (2/3)c s_2(r)$, $s_2(r)$ is the solution of~\eqref{eq:diffeq}, and $r_0$ is the smallest positive root of $s_2(r) = 0$.
\end{lemma}

\noindent
{\bf Proof.}
Given Lemma~\ref{lem:round} and including the $o(1)$ probability that the state is not within $o(n)$ of that predicted by the differential equations for all $r$, we can write
\begin{eqnarray*}
 P_c(0) & = & \left( \prod_R q_\success(R/n) \right) + o(1) \\
 & = & \prod_R \exp\left( -\frac{f(\lambda(r))}{n} + o(1/n) \right) + o(1) \\
 %\left( \prod_R q_\success(R/n) \right) + o(1) \nonumber \\
 & = & \exp\left(- \frac{1}{n} \sum_R f(\lambda(R/n)) \right) + o(1) \\
 & = & \exp\left(- \int_0^{r_0} \dr f(\lambda(r)) \right) + o(1) \enspace . 
\end{eqnarray*}
In the second line we used $\ln (1-x) = -x + O(x^2)$, and in the last line we used the fact that $f(\lambda(r))$ is bounded and differentiable as long as $\lambda(r) < 1$.
\qed

Finally, we obtain~\eqref{eq:p0t} from~\eqref{eq:p0} by changing the variable of integration from $r$ to $t$.  Since $\dt/\dr = 1/(1-\lambda)$, this gives
\begin{equation}
%\label{eq:p0t}
 P_c(0) = \exp\left( - \int_0^{t_0} \dt \,\frac{c \lambda^2}{2 (1-\lambda)(2+\lambda)} \right) + o(1) 
 \enspace .
\end{equation}
Here $\lambda = (2/3)cs_2(t)$ and $t_0$ is the time at which we complete the giant component, or equivalently, the first time after we start coloring the giant component at which the number of 2-color vertices becomes zero.  Using~\eqref{eq:sol}, this is the smallest positive root of
\begin{equation}
\label{eq:t0}
 s_2(t) = 1 - t - \e^{-ct} = 0 
\end{equation}
completing the proof of Theorem~\ref{thm:p0}.  
%More explicitly, \begin{equation}
%\label{eq:t0}
% t_0 = 1 + \frac{1}{c} W(-c\e^{-c}) \enspace .
%\end{equation}
\qed

\begin{figure}
\begin{center}
\includegraphics[width = 3.5in]{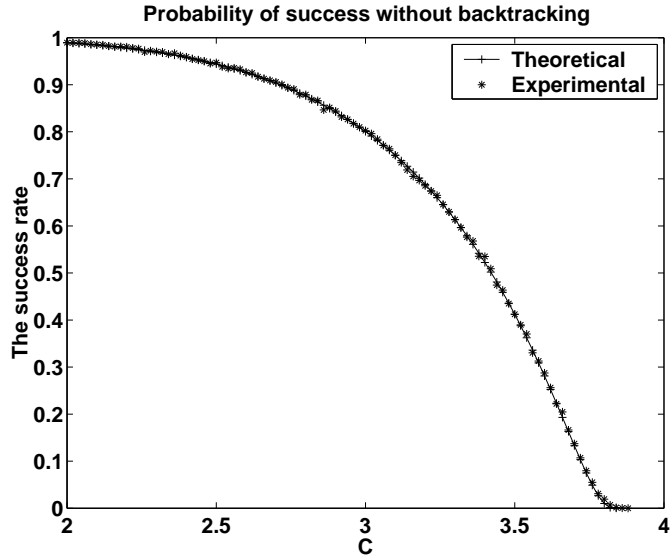}
\end{center}
\caption{A comparison of our calculation~\eqref{eq:p0t} of the probability $P_c(0)$ of success without backtracking (the solid line) with experimental results (the stars) as a function of $c$.  The experiments consisted of $10^4$ trials for each value of $c$, on graphs of size $n=10^4$.}
\label{fig:p0}
\end{figure}

We have not found a closed form for the integral in~\eqref{eq:p0t}.  However, Figure~\ref{fig:p0} compares values of $P_c(0)$ obtained by integrating~\eqref{eq:p0t} numerically with experimental data for graphs of size $n=10^4$, and they are in excellent agreement.

\subsection{Another approach: the distribution of 1-color vertices}
\label{sec:heuristic}

In this section, we look at a heuristic calculation of $P_c(0)$ in which we consider the steps of \gl\ one at a time, rather than in rounds.  This method is analytically simpler than that in the previous section.  However, to make it rigorous, we would need to deal with the fact that the events that a pair of nearby steps fail are positively correlated if they occur in the same round; for instance, they are both more likely to fail if that round colors many vertices.  (One way to remove this correlation would be to note that the probability that more than one step in a given round fails is $O(\polylog(n)/n^2)$, so taking a union bound over the $O(n)$ steps, with probability $1-o(1)$ no two steps fail in the same round.)

We start by calculating the probability distribution $p(x)$ of the number of 1-color vertices that are present at a given time, since the probability that two of these are neighbors is essentially $p$ times its second moment.  If we think of \gl\ in single steps rather than in rounds, Achlioptas and Molloy~\cite{AchMolloy} showed that $x$ obeys a biased random walk, where at each step we first decrement $x$ if it is positive (since we color a 1-color vertex if one exists) and then increase it by a random variable $y$ which is Poisson-distributed with mean $\lambda$ (since we create $y$ new 1-color vertices).  

Since $\lambda$ varies continuously with $t$, as $n \to \infty$ we can assume that $\lambda$ is roughly constant over a large number of steps, in which case $p(x)$ will be close to the stationary distribution of this biased random walk.  We can calculate $p(x)$ using its generating function
\[
 g(z) =  \sum_{x=0}^{\infty}  p(x)  z^x
\]
In particular, $g(0) = p(0)$ is the probability that there is no 1-color vertex, i.e., that the current step is a free step.  Since the expected change in $x$, which is $p(0) - 1 + \lambda$, must be zero for the stationary distribution, we have $p(0) = 1-\lambda$.
% and $g(1) = \sum_{x=0}^{\infty} p(x) = 1$. 
%\begin{eqnarray*}
% \frac{\dg(z))}{\dz} & =  &\sum_{x=0}^{\infty} p(x) x z^{x-1} \\ 
%& = & \sum_{x=0}^{\infty} p(x)  x  \\
%&  =  & \overline{x} \normalsize \ \ \ \ \ \ at \   z\ =\ 1 
%\end{eqnarray*}

Decrementing $x$ by $1$ if  $x > 0$ corresponds to dividing $g(z)$ by $z$ except for the $z^0$ term, and adding $y$ to $x$ corresponds to multiplying $g(z)$ by the generating function of the Poisson distribution, 
\[ 
 \sum_{y=0}^{\infty} \frac{\e^{-\lambda} \lambda^y }{y!} z^y 
   = \e^{\lambda(z-1)} \enspace .
\]
Thus the effect of each step on $g(z)$ is
\[ g(z) \mapsto \left( \frac{g(z)-p(0)}{z} + p(0) \right) \e^{\lambda\left( z-1 \right)} 
= \bigl( g(z) + (1 - \lambda)(z-1) \bigr) \frac{\e^{\lambda\left( z-1 \right)}}{z}  \]
and solving for the stationary distribution gives
\[ g(z) 
%= \frac{ p(0)  \left( z -1 \right) } {z \e^{-\lambda(z-1)} - 1} 
          =  \frac{ \left( 1-\lambda \right) \left( z -1 \right) } {z \e^{-\lambda(z-1)} - 1} \enspace . \]
The expected number of 1-color vertices present on a given step is then
\[ \ex{x} = g'(1) = \frac{ \lambda (2 - \lambda) }{2 (1 - \lambda)} \enspace . \]
and the expected number of 1-color vertices other than the one colored on a given step is
\begin{equation}
\label{eq:other}
 \ex{x}-1+p_0 = \ex{x} - \lambda = \frac{{ \lambda}^2}{2 (1 - \lambda)} 
\end{equation}
any of which could conceivably become a $0$-color vertex on that step.

However, the colors of the existing 1-color vertices are correlated with each other, so we can't simply divide~\eqref{eq:other} by $3$.  Thinking back to the tree $T$ generated by a round, if two colors are $k$ steps apart in the tree, then the probability that they are the same color is given by
\[ F(k) = \frac{1- F(k-1)}{2} 
  = \frac{1}{3}  \left( 1 + 2  \left(- \frac{1}{2} \right)^{\!k\,} \right) 
\]
e.g. $F(0) = 1$, $F(1) = 0$ (since edges in $T$ only connect 1-color vertices of different colors), $F(2) = 1/2$, and so on.

In a branching process of branching ratio $\lambda$, the average number of vertices $k$ steps away from a given vertex is $\lambda^k$.  Summing over all $k \ge 1$ and dividing by $2$ since we are counting each pair of vertices twice, the expected number of partners forming a dangerous pair with the vertex colored on a given step is
\begin{eqnarray*}
  \frac{1}{2} \sum_{k=1}^{\infty}  \lambda^k \, F(k)  
  & = & \frac{1}{6} \sum_{k=1}^{\infty}  \lambda^k 
    + \frac{1}{3} \sum_{k=1}^{\infty} \left(- \frac{\lambda}{2}\right)^{\!k\,} \\ 
  & = & \frac{1}{6} \frac{\lambda}{1-\lambda} - \frac{1}{3} \frac{\lambda}{2+  \lambda} \\
  & = & \frac{\lambda^2}{2(1-\lambda)(2+\lambda)} \enspace . 
\end{eqnarray*}
Multiplying this by $p=c/n$ and integrating over the steps $0 < t < t_0$ gives the same integral for the expected number of $0$-color vertices created while coloring the giant component as in~\eqref{eq:p0t}.

\subsection{The singularity at $c^*$ and the difficulty of numerical experiments}
\label{sec:singularity}

As $c$ approaches $c^*$, the maximum value of $\lambda$ approaches $1$, and the integral in~\eqref{eq:p0t} diverges.  To isolate the nature of this divergence, we expand the integrand in terms of partial fractions, which gives
\begin{equation}
\label{eq:partial}
 - \ln P_c(0) =  \frac{c}{6} \int_0^{t_0} \dt \left( \frac{1}{1-\lambda} - \frac{2+3\lambda}{2+\lambda} \right) 
  = \frac{c}{6} \int_0^{t_0} \frac{\dt}{1-\lambda} - O(1) \enspace .
\end{equation}
Given~\eqref{eq:diffeqt} and~\eqref{eq:sol}, $s_2$ and $\lambda$ are maximized at $\tmax = (\ln c)/c$.  Expanding $\lambda$ as a Taylor series in $t$ around $\tmax$ gives
\begin{equation}
\label{eq:taylor}
 \int_0^{t_0} \frac{\dt}{1-\lambda}
\approx \int \frac{\dt}{1-\lmax-(1/2) \lambda'' (t-\tmax)^2}
= \frac{\pi}{\sqrt{1-\lmax} \sqrt{-\lambda''/2}}
\end{equation}
where
\[ \lambda'' = -\frac{2}{3} c^2 \]
and
\[ \lmax = \frac{2}{3} (c - \ln c - 1) \]
Let $c = c^* - \eps$ where $c^*$ is the unique positive root of $c - \ln c = 5/2$~\cite{AchMolloy}.  To leading order in $\eps$, we have
\[ 1 - \lmax \approx \eps \left. \frac{\p \lmax}{\p c} \right|_{c=c^*} = \frac{2(c^*-1)}{3c^*} \,\eps \]
and~\eqref{eq:partial} and~\eqref{eq:taylor} give
\[ - \ln P_c(0) = \frac{A}{\sqrt{\eps}} - O(1) \]
where
\[ A = \frac{\pi}{2} \sqrt{\frac{c^*}{2(c^*-1)}} \approx 1.29 \enspace . \]
Thus the probability of success is given by
\[ \lim_{\eps \to 0} P_c(0) = \exp(-A/\sqrt{\eps}) \,\Theta(1) \]
which goes to zero faster than any analytic function as $\eps \to 0$.  In particular, all of its derivatives with respect to $c$ are zero at $c^*$.

While the threshold $c^*$ below which \gl\ succeeds with positive probability can be determined analytically, more sophisticated heuristics often require numerical experiments --- if only to confirm a long and involved journey through a large system of coupled differential equations.  However, since $P_c(0)$ approaches zero very rapidly as $c \to c^*$, the number of trials we have to do to confirm that \gl\ succeeds with positive probability increases very rapidly.  
Using methods from statistical physics, Deroulers and Monasson~\cite{DeroulersMonasson} 
found the same critical behavior for heuristics on random 3-SAT;
we expect a similar pattern for other heuristics, such as the smoothed Brelaz 
heuristic analyzed by Achlioptas and Moore~\cite{AchMoore} which 
succeeds for $c < 4.03$.

To illustrate this, in Table~\ref{tab} we show $P_c(0)$ for various values of $c$.  Note that to measure $c^*$ to one, two, or three decimal digits, we need to do roughly $10^2$, $10^6$, and $10^{28}$ trials!  On a practical level, this means that numerical experiments will systematically underestimate the threshold below which a heuristic of this type succeeds with positive probability.
\begin{table}[ht]
\centerline{ $
\begin{array}{c|c|c}
c & -\ln P_c(0) & P_c(0) \\ \hline
3.8 & 4.569 & 0.0104 \\
3.84 & 13.654 & 1.176 \times 10^{-6} \\
3.847 & 63.467 & 2.733 \times 10^{-28}
\end{array} 
$ }
\caption{The rapid decrease of $P_c(0)$ as $c$ approaches $c^* \approx 3.8474$.}
\label{tab}
\end{table}

\section{Exponential backtracking with positive probability}
\label{sec:exp}

In this section we prove Theorem~\ref{thm:main}, establishing rigorously that the number of backtracks of DPLL on random graphs with degree $1 < c < c^*$ has a heavy tail.
\medskip

\noindent
{\bf Proof of Theorem~\ref{thm:main}.}  We focus on algorithm \ord\ first, in which each vertex is given an index in a fixed random order.  Let $t_1$ be a constant such that $0 < t_1 < t_0$ where $t_0$ is given by~\eqref{eq:t0}.  Run the algorithm for $t_1 n$ steps, and then continue until the end of the current round (which takes with high probability $o(n)$ more steps), conditioning on not having created a 0-color vertex so far.  This is equivalent to running \gl\ conditioned on its success, so as discussed above, at the end of these $t_1 n + o(n)$ steps there are with high probability  $s_3(t_1)n+o(n)$ 3-color vertices and $s_2(t_1)/3 + o(n)$ 2-color vertices of each color pair, where $s_3(t_1)$ and $s_2(t_1)$ are given by~\eqref{eq:sol}.  In addition, the uncolored part of the graph $G'$ is uniformly random in $G(n',p)$ where $n'$ is the total number of uncolored vertices.  

Let us call a triangle {\em bad} if it is composed of 2-color vertices whose allowed colors are red and green, it is disconnected from the rest of $G'$, and the indices of its vertices are all greater than the median index of the 2-color vertices in $G'$.  Now, let $E_1$ be the event that $G'$ contains exactly one bad triangle.  It is easy to see that the distribution of the number of bad triangles is within $o(1)$ of a  Poisson distribution with expectation 
\[ m = \frac{1}{8} {s_2(t_1) n/3 \choose 3} p^3 (1-p)^{3s_3(t_1)n}
= \frac{c^3 s_2(t_1)^3 \,\e^{-3 c s_3(t_1)}}{1296} = \Theta(1) \]
Then $E_1$ occurs with probability $q_1 = m \e^{-m} + o(1) > 0$. 

Let us call this triangle $\Delta$.  It is important to us in the following ways:
\begin{enumerate}
\item It is not 2-colorable, so every branch of this subtree will fail, and the algorithm will be forced to backtrack at least to the $(t_1 n)$th step and uncolor one of $\Delta$'s blue neighbors.
\item Since $\Delta$ is isolated from rest of $G'$, we will find this contradiction only if we choose one of
$\Delta$'s vertices from the pool of 2-color vertices; we will not be led to $\Delta$ by a chain of forced steps.
\item When running \ord, we won't choose any of $\Delta$'s vertices until we run out of 2-color vertices of lower index, and this will not happen until we have taken at least $s_2(t_1)n/2$ more steps.
 \end{enumerate}   
 In other words, $\Delta$ will cause the entire subtree starting with these $t_1 n$ steps to fail, but we won't find out about it until we explore the tree $\Theta(n)$ more deeply, and visit an exponential number of nodes.
 
To formalize this, let $t_2$ be a constant such that $0 < t_2 < s_2(t_1)/2$, and consider running the algorithm for another $t_2 n$ steps.
%, with the modification that if a $0$-color vertex is created, we continue branching on other 2-color vertices instead of backtracking.  
This produces a search tree of depth $t_2 n$, where each internal node corresponding to a forced or free step has one or two children respectively.  If we choose a random branch of this tree by following the two branches with equal probability each time we come to a free step, this is equivalent to running \gl\ on the graph $G'' = G' \setminus \Delta$.  Each leaf of the tree corresponds either to creating a $0$-color vertex and backtracking, or having run for $t_2 n$ steps without creating a $0$-color vertex.  We call these ``bad'' and ``good'' leaves respectively.

We will abuse notation by letting $G(n'-3,p)$ denote a random graph with three fewer red-green vertices than $G'$.  Once we condition on the number of uncolored vertices of each color list in $G'$ and on the event that $E_1$ occurs, $G''$ is uniformly random in $G(n'-3,p)$ except for the condition that it has no bad triangles (it is easy to see this given the structure of $G(n,p)$ as a product space).  The progress of \gl\ on $G(n'-3,p)$ is still given by the differential equations~\eqref{eq:diffeqt}, since removing three vertices changes the rescaled variables by $o(1)$.  Since there is one free step per round, the number of free steps performed by $t_2 n$ steps of \gl\ on $G(n'-3,p)$ is with high probability  $\alpha n + o(n)$ where $\alpha = r(t_2)-r(t_1)$ and $r(t)$ is given by~\eqref{eq:r}.  But, the event that $G(n'-3,p)$ has no bad triangles occurs with probability $\e^{-m}+o(1) = \Theta(1)$, so conditioning on this event the number of free steps performed by \gl\ on $G''$ is still with high probability $\alpha n + o(n)$.  Furthermore, \gl\ succeeds on $G(n'-3,p)$ for $t_2 n$ steps with probability at least $P_c(0)$, and since this success implies the condition that $G(n'-3,p)$ has no bad triangles, \gl\ succeeds on $G''$ with probability $P \ge P_c(0)$.

Let us transform the search tree to a binary tree $T$ with the same number of leaves, by replacing each chain of forced steps with a single edge, and leaving just the internal nodes corresponding to free steps.  The depth of a leaf is now the number of free steps on the way to it, and a run of \gl\ samples a given leaf at depth $i$ with probability $2^{-i}$.  Let $M$ be the average depth of a good leaf according to this probability distribution; then with high probability $M=\alpha n + o(n)$, and the total probability of the good leaves is $P$.

We wish to prove a lower bound on the number of leaves.  If $T$ were perfectly balanced, this would be easy; but unfortunately $M$ is not exponentially concentrated, so the depth of the leaves can vary significantly.  Therefore, we employ the following lemma.

\begin{lemma}  \label{lem:tree}
Let $T$ be a binary tree.  Assign a probability $2^{-i}$ to each leaf at depth $i$, and label each leaf ``good'' or ``bad.''  Let $M$ be the average depth of the good leaves, and let $P$ be their total probability.  Then there are at least $P \,2^M$ good leaves.
\end{lemma}

\noindent
{\bf Proof of Lemma~\ref{lem:tree}.}
Let $N$ be the number of good leaves; we prove the lemma by induction on the size of the tree.  For the base case, a tree consisting of a single vertex has $P=1$, $M=0$ and $N=1$ if it is good, and $P=0$ and $N=0$ if it is bad.

Now assume inductively that the lemma is true for $T$'s subtrees.  Let $N_\ell$ and $N_r$ denote the number of good leaves of the left and right subtrees, $M_\ell$ and $M_r$ their average depth (measured from the subtrees' roots) and $P_\ell$ and $P_r$ their conditional probabilities.  Then we have $N = N_\ell + N_r$, $P=(P_\ell + P_r)/2$, and 
\[ M = \frac{P_\ell M_\ell + P_r M_r}{2P} + 1 \enspace . \]
Let $p,q \ge 0$ and $p+q = 1$.  Then for any $A, B \ge 0$ we have
\begin{equation}
\label{eq:mean}
  p A + q B \geq A^p B^q 
\end{equation}
i.e., the weighted arithmetic mean is at least as large as the weighted geometric mean.  Then taking $p = P_\ell/(2P)$ and $q = P_r/(2P)$, we have
\begin{eqnarray*}
N & = & N_\ell + N_r \\
& \geq & P_\ell \,2^{M_\ell} + P_r \,2^{M_r} \\
& \geq & (2P) \,2^{(P_\ell M_\ell + P_r M_r)/(2P)} \\
& = & P \,2^M \enspace . 
\end{eqnarray*}
\qed

%For an alternate proof, let $n(i)$ be the number of good leaves at depth $i$.  Since $\sum_i n(i) \,2^{-i} = P$, we can think of $(1/P) \,n(i) \,2^{-i}$ as a probability distribution on the depth.  In that case $N=\sum_i n(i) = P \cdot \ex{2^i}$ and $M = (1/P) \sum_i n(i) \,2^{-i} i = \ex{i}$, and we have $\ex{2^i} \geq 2^{\ex{i}}$ by Jensen's inequality.

Lemma~\ref{lem:tree} and the above arguments imply that with probability $q_1 - o(1)$, \ord\ will backtrack at least $P_c(0) \,2^{\alpha n - o(n)} = 2^{\alpha n - o(n)}$ times.  Taking any $q < q_1$ and any $\beta < \alpha$ completes the proof for \ord.
\smallskip

The proof for algorithm \rand\ is similar.  We remove the condition on $\Delta$'s indices (removing the factor of $1/8$ from $m$ above).  However, the branches of $T$ can now fail either because 
\begin{enumerate}
\item \gl\ creates a $0$-color vertex while running on $G''$, or 
\item the algorithm chooses to branch on one of $\Delta$'s vertices.  
\end{enumerate}
Let $\smin = \min_{t_1 \le t \le t_2} s_2(t)$.  Then the probability that one of $\Delta$'s vertices is chosen on a given step is with high probability at most $3/(\smin n + o(n))$, and the probability a branch fails for the second reason is at most $3 t_2 / \smin + o(1)$.  The probability of the good leaves is then $P \ge P_c(0) - 3 t_2 / \smin - o(1)$, and by taking $t_2$ sufficiently small we can ensure that $P > 0$.  Thus \rand\ also backtracks $2^{\alpha n - o(n)}$ times with probability $q_1 - o(1)$.  We again take $q < q_1$ and $\beta < \alpha$, and the proof is complete.

\section{Discussion}
\label{sec:dis}

We have shown that DPLL algorithms take exponential time with positive probability for random graphs $G(n,c/n)$, even in the ``easy'' range $1 < c < 3.847...$ where with positive probability they color the graph with no backtracking at all.  This happens because the events that different branches of the search tree fail are far from independent; since a single bad triangle $\Delta$ dooms an entire subtree to failure, the probability all its branches fail is positive even though a random branch succeeds with positive probability.  The algorithm then tries to 2-color $\Delta$ an exponential number of times, naively hoping that recoloring other vertices will render $\Delta$ 2-colorable.  In terms of restarts, once $\Delta$ has ``spoiled'' an entire section of the search space, it makes more sense to start over with a new random branch. 

Experimentally, Figure~\ref{fig:c35} shows that  the distribution of the number of backtracks follows a power law $P_c(b) \sim b^{-1}$.  It might be possible to strengthen Theorem~\ref{thm:main} to prove this power-law behavior in the following way: suppose for the sake of argument that $\Delta$ appears at a uniformly random depth $d$ between $1$ and $n$, and that the running time $b$ is exactly $2^{Ad}$ for some $A$.  Then the probability that $b$ is between $2^{Ad}$ and $2^{A(d+1)}$ is $1/n$, giving a probability density $P_c(b) = 1/(2^{Ad} (2^A-1) n) \sim 1/b$.  Of course, $d$ is not uniformly distributed, but any distribution which varies slowly from $\Theta(1)$ to $\Theta(n)$ would give the same qualitative result.  The difficulty is determining how $d$ is distributed, and then better understanding the distribution of $b$: however, bounding $b$'s variance, say, seems quite challenging.  We propose this as a direction for future work.

 \section*{Acknowledgments}
 
We are grateful to Dimitris Achlioptas, Sinan Al-Saffar, Paul Beame, Tracy Conrad, Michael Molloy, Remi Monasson, Bart Selman and Vishal Sanwalani for helpful comments and conversations.  This work was supported by NSF grant PHY-0200909 and the Los Alamos National Laboratory.
H.J. is supported by a NSF Graduate Research Fellowship.

\end{document}